\documentclass[prl,showpacs,byrevtex,floatfix,twocolumn,superscriptaddress]{revtex4}
\usepackage{graphicx}
\usepackage{dcolumn}
\usepackage{bm}

\begin{document}

\bibliographystyle{prsty}

\preprint{Draft - not for distribution}

\title[Edge]{\boldmath
  The absence of superfluid response in $ac$ and $bc$-plane optical
  conductivities of optimally-doped Bi$_2$Sr$_2$CaCu$_2$O$_{8+\delta}$ single
  crystals in the surface region
  \unboldmath }
\author{J. J. Tu}
\affiliation{Department of Physics, City College of the City University of New
York, New York, New York 10031}
\author{C. C. Homes}
\email{homes@bnl.gov}%
\affiliation{Department of Physics, Brookhaven National Laboratory, Upton, New
York 11973}
\author{L. H. Greene}
\affiliation{Department of Physics and the Fredrick Seitz Materials Research
Laboratory, University of Illinois at Urbana-Champaign, Urbana, Illinois 61801}
%
%
%
\author{G. D. Gu}
\author{M. Strongin}%
\affiliation{Department of Physics, Brookhaven National Laboratory, Upton, New
York 11973}
\date{\today}

%
%
\begin{abstract}
The optical properties of optimally-doped Bi$_2$Sr$_2$CaCu$_2$O$_{8+\delta}$
(Bi2212) have been measured normal to the edge planes [{\it ac} plane, {\it bc}
plane, and $(1\bar{1}0)$ plane], for light polarized parallel to nodal and
anti-nodal (gap) directions, respectively.  While the superfluid contribution
can be obtained from the optical conductivities in the $(1\bar{1}0)$-plane, it
is unobservable in the {\it ac} and {\it bc}-planes. This apparent asymmetry
implies that the edge region of high-T$_c$ cuprates is unusual and further
supports a {\it d}-wave symmetry of the superconducting order parameter.
\end{abstract}

%

%
%
\pacs{74.25.Gz, 74.72.Hs, 74.25.Nf}

\maketitle

%
%
Understanding the microscopic mechanism of high-temperature (high-T$_c$)
superconductors remains one of the fundamental challenges of condensed matter
physics. Phase-sensitive techniques \cite{tsuei94, wollman95,harlingen95,
covington97,sutherland03} imply a {\it d}-wave symmetry of the superconducting
order parameter. Angle-resolved photoemission (ARPES) results in Bi2212
\cite{valla99} clearly show an aniostropic energy gap.
However, {\it ab}-plane optical conductivity measurements on the same material
do not show large anisotropies with light (electric field vector) polarized
parallel to {\it a}-axis, {\it b}-axis, or anti-nodal directions
\cite{quijada99,tu02a} even below T$_c$. This can be understood since ARPES is
a {\it k}-dependent measurement, whereas the optical conductivity is averaged
over the Fermi surface. In general, one would expect the same kind of optical
results for the edge planes [the {\it ac}-plane, {\it bc}-plane, and
$(1\bar{1}0)$-plane]. These edge regions of high-T$_c$ superconductors were
studied extensively by tunneling experiments on YBa$_2$Cu$_3$O$_{6+x}$
\cite{geerk88,lesueur92,covington97,wei98,alff98,krupke99} and more recently on
Bi2212 by Greene and co-workers \cite{aubin02}. One particularly intriguing
feature is the zero bias conductance peak (ZBCP) observed on the {\it ac} or
{\it bc}-faces, coupled with the absence of a gap feature when tunneling into
{\it ab}-plane. This is contrasted with the observation of a weaker ZBCP when
tunneling into the $(1\bar{1}0)$-plane, with the appearance of a
superconducting gap \cite{aubin02}.  We emphasize that for this work we take
the crystallographic {\it a} and {\it b} axis along the Cu-Cu bonds as the
nodal direction, whereas the $[1\bar{1}0]$ direction (the anti-nodal or
gap-maximum direction) is along the Cu-O bonds.

In the superconducting state a particle bound state forms at the Fermi surface
when the node of a {\it d}-wave order parameter is normal to a reflecting
surface \cite{hu94}, such as the {\it ac} and {\it bc}-faces of Bi2212.
Particles reflecting from such surfaces experience a change in the sign of the
order parameter along their classical trajectory and subsequently undergo
Andreev reflection. Constructive interference between incident and
Andreev-reflected particles leads to the formation of bound states confined to
the surface. These bound states will produce a ZBCP in a tunneling spectrum
\cite{tanaka95,fogelstrom97,carrington01}.  Andreev scattering causes strong
pair breaking, which leaves a surface region depleted of superfluid. The
motivation of this study is to examine systematically these surface regions in
Bi2212 by measuring the optical conductivities. The question is whether the
picture used to explain the tunneling measurements, which probe a surface
region of the order of $\simeq 10$~nm, can be used to describe the wider
surface region probed by infrared radiation, which is typically $\simeq
100$~nm.

In this Letter, we report characteristically different behavior observed in the
{\it ac} and {\it bc}-plane conductivities of optimally doped Bi2212 as
compared to the $(1\bar{1}0)$-plane conductivities below T$_c$.  While the
superfluid contribution can be measured in the optical conductivities in the
$(1\bar{1}0)$-plane, it is much smaller in the {\it ac} and {\it bc}-plane.
This apparent asymmetry implies that the edge region in high-T$_c$ {\it d}-wave
superconductors has unusual properties that are different from the bulk.

%
%
The {\it ab}-plane optical conductivity of optimally-doped Bi2212 has been
measured extensively \cite{reedyk88,puchkov96,quijada99,wang99,tu02a}. However,
because of the large {\it c}-axis dimension required to carry out optical
measurements on the {\it ac}, {\it bc} and $(1\bar{1}0)$ faces, only one brief
study was previously reported \cite{tajima93}.  For this study, large
optimally-doped Bi2212 single crystals are grown using the
traveling-surface-floating-zone (TSFZ) method.  The typical size of these
crystals for the edge experiments is $5\times 3 \times 1$~mm$^3$ along the
three principle crystallographic axes. Cleaved (001) surfaces are used for the
{\it ab}-plane measurements. However, to study the edge regions, polished
(100), (010) and $(1\bar{1}0)$ surfaces are required.
Considerable care has been taken during polishing due to the mica-like nature
of Bi2212.  Polishing has been done by hand, and always along the planar
direction.  A final polish with 0.1~$\mu$m diamond films allows optical surface
quality to be achieved.
The surface quality of our polished samples should be comparable to that of the
samples used in the tunneling experiments on Bi2212 \cite{aubin02}, which have
been found to have a surface roughness of $\approx 80$~\AA\ measured by AFM.
The Bi2212 crystals are mounted on an optically-black cone, and the
temperature-dependent polarized reflectance is measured in a
near-normal-incidence arrangement from $\approx 50$ to over 16,000 cm$^{-1}$ on
a Bruker IFS~66v/S. The absolute reflectivity is determined by evaporating a
gold film {\it in situ} over the sample \cite{homes93}. This comparison to the
gold reflectivity provides an absolute reflectivity scale. The optical
conductivities are then determined from a Kramers-Kronig analysis.

%
%
\begin{figure}[t]
\vspace*{-0.5cm}%
%
%
\centerline{\includegraphics[width=3.0in]{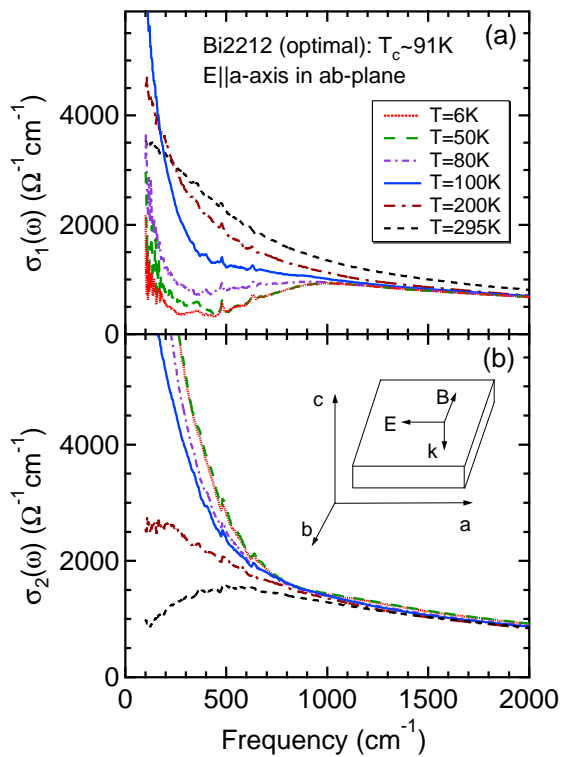}}%
\vspace*{-0.3cm}%
\caption{The {\it ab}-plane conductivity data of an optimally-doped Bi2212
single crystal for ${\rm E}\!\parallel\!{a}$. (a) Temperature-dependent
$\sigma_1$; (b)temperature-dependent $\sigma_2$.  Inset: the experimental
configuration.}%
\vspace*{-0.4cm}%
\label{fig:ab}
\end{figure}

The temperature-dependent {\it ab}-plane conductivity data is shown in
Fig.~\ref{fig:ab} for a single-crystal Bi2212 sample with for light polarized
along the {\it a} axis (${\rm E}\!\parallel\!{a}$).  In agreement with the
previous results \cite{quijada99}, there is only a weak dependence of the
conductivity on the direction of the polarization within the {\it ab}-plane.
However, strong phonon anisotropy has been observed in our {\it ab}-plane
conductivity measurements \cite{tu02a,tu02b}. The superfluid response is
observed in the {\it ab}-plane conductivities below T$_c$, as $\sigma_1$
decreases with temperature according to the Ferrell-Glover-Tinkham sum rule
accompanied by a simultaneous increase in $\sigma_2$. The {\it ab}-plane data
is presented here as a reference to show the large difference from the edge
plane data presented in the next figure.

%
%
\begin{figure}[t]
\vspace*{-0.5cm}%
%
%
\centerline{\includegraphics[width=3.0in]{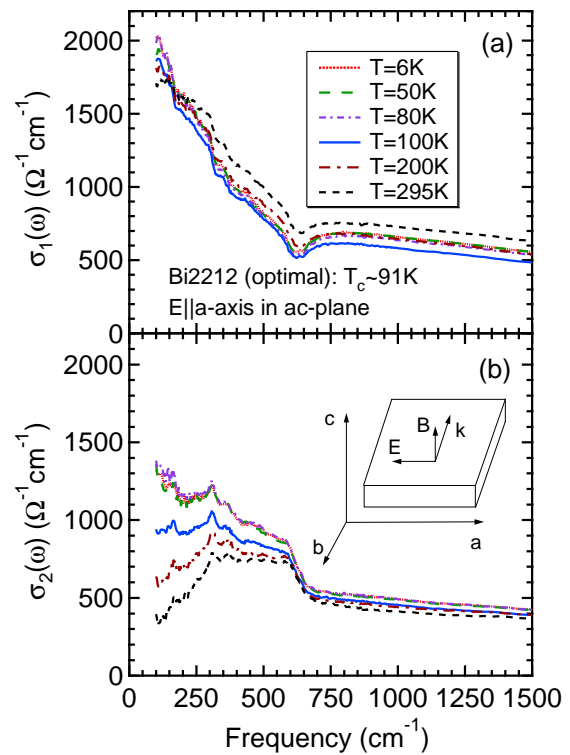}}%
\vspace*{-0.3cm}%
\caption{The {\it ac}-plane conductivity data of an optimally-doped Bi2212
single crystal for ${\rm E}\!\parallel\!{a}$. No superfluid response is
observed in either $\sigma_1$ or $\sigma_2$. (a) Temperature-dependent
$\sigma_1$; (b)temperature-dependent $\sigma_2$.
Inset: the experimental configuration.}%
\vspace*{-0.4cm}%
\label{fig:ac}
\end{figure}

\begin{figure}[t]
\vspace*{-0.5cm}%
%
%
\centerline{\includegraphics[width=3.0in]{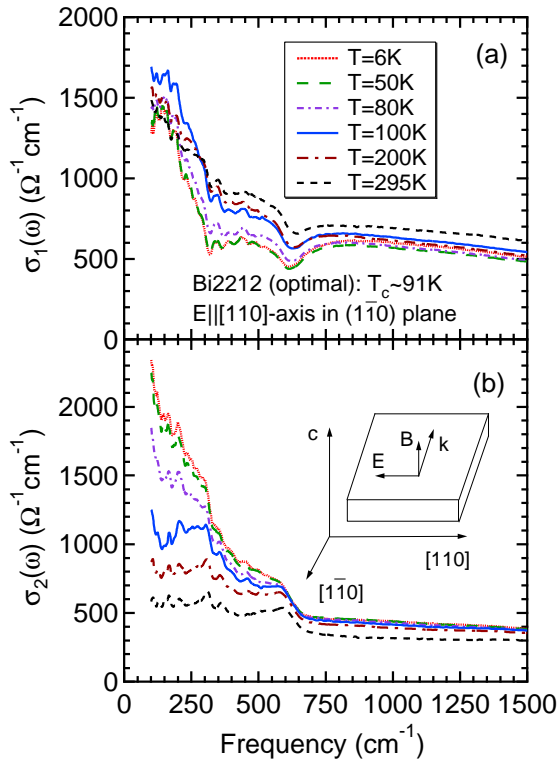}}%
\vspace*{-0.3cm}%
\caption{The $(1\bar{1}0)$-plane conductivity data of an optimally-doped Bi2212
single crystal for ${\rm E}\!\parallel\![110]$. In contrast to {\it ac}-plane
conductivities, the superfluid response is observed in $\sigma_1$ and
$\sigma_2$ in this case. (a) Temperature-dependent $\sigma_1$;
(b) temperature-dependent $\sigma_2$.  Inset: the experimental configuration.}%
\vspace*{-0.4cm}%
\label{fig:nodal}
\end{figure}

The temperature-dependent {\it ac}-plane conductivity data is shown in
Fig.~\ref{fig:ac} for a single-crystal Bi2212 sample for ${\rm
E}\!\parallel\!{a}$.  The temperature-dependent conductivity data measured on a
$(1\bar{1}0)$-plane is shown in Fig.~\ref{fig:nodal} for ${\rm E}\!\parallel\!
[110]$. The main point of comparing the {\it ac}-plane and the
$(1\bar{1}0)$-plane conductivity, is that while the superfluid contribution to
optical conductivity is observed in the $(1\bar{1}0)$-plane as shown in
Fig.~\ref{fig:nodal}, it is unobservable in {\it ac}-plane as given in
Fig.~\ref{fig:ac}.  In both cases, $\sigma_1$ at room temperature is similar,
but significantly lower than the {\it ab}-plane value. However, it can be seen
that as the temperature is lowered below T$_c$, which is $\approx 91$~K for
these optimally-doped samples, the behavior of the $(1\bar{1}0)$-plane
conductivity data is much closer to the {\it ab}-plane data, showing the
characteristic decrease of $\sigma_1$ as normal carriers start to condense into
superfluid which leads to a significant increase of $\sigma_2$ below T$_c$. The
behavior of the {\it ac}-plane conductivity around and below T$_c$ is very
different. As the temperature changed from 100 to 80~K, both $\sigma_1$ and
$\sigma_2$ show a significant increase, particularly in $\sigma_2$ below 500
cm$^{-1}$ [Fig.~\ref{fig:ac}(b)].  As the temperature is lowered further, no
noticeable changes are observed in $\sigma_1$ or $\sigma_2$. Similar results
are obtained for the {\it bc}-plane conductivity as compared to the {\it
ac}-plane conductivity.
%

%
%

The essence of the our results is the difference between Fig.~\ref{fig:ac} and
Fig.~\ref{fig:nodal}, where the conductivities in the surface regions are
compared for a surface with a normal along a nodal direction and a surface with
a normal along an anti-nodal direction.  It immediately shows that while the
superfluid contribution to optical conductivity can be observed in the surface
region of the $(1\bar{1}0)$-plane, it is unobservable in the surface region of
the {\it ac} or {\it bc} planes. This apparent asymmetry implies that the
surface region in the high-T$_c$ {\it d}-wave superconductors has unusual
properties that are different from the bulk \cite{muller03}.
%
%

%
%
\begin{figure}[t]
\vspace*{-0.5cm}%
%
%
%
%
\centerline{\includegraphics[width=3.0in]{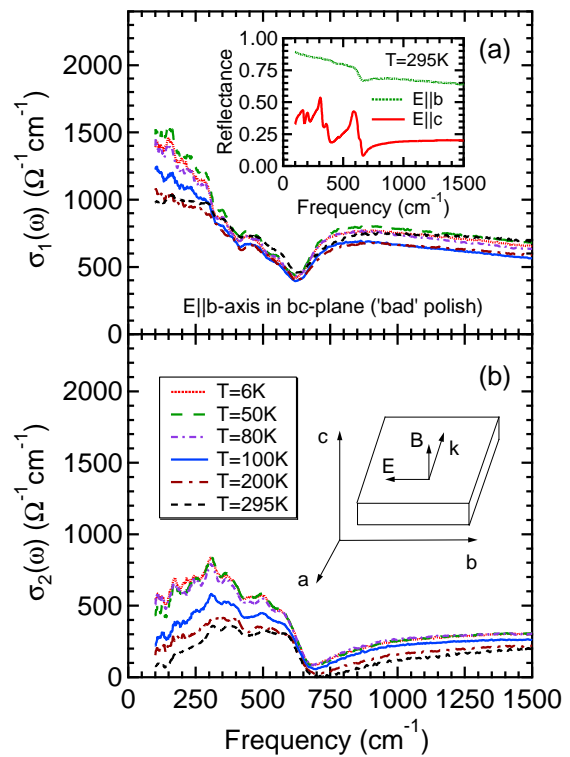}}%
\vspace*{-0.3cm}%
\caption{The {\it bc}-plane conductivity data of an optimally doped Bi2212
single crystal that has a coarser surface finish for ${\rm E}\!\parallel\!{b}$.
No superfluid response is observed in either $\sigma_1$ or $\sigma_2$. (a)
Temperature-dependent $\sigma_1$. Inset: the room-temperature reflectance data
for ${\rm E}\!\parallel\!{b}$ and ${\rm E}\!\parallel\!{c}$, respectively. (b)
Temperature-dependent $\sigma_2$. Inset: the experimental configuration.}%
\vspace*{-0.4cm}%
\label{fig:polish}
\end{figure}

The comparison with the {\it ab}-plane conductivities shows that the
conductivities in these edge regions at room temperature are reduced by about a
factor of two, and this is no doubt due to the polishing process.  To further
understand the role of disorder induced by the polishing process, we have added
a set of data to show that with a coarser polishing finish ($1\,\mu$m diamond
film) the conductivity of the {\it bc}-plane in the normal state is even more
drastically reduced, as shown in Fig.~\ref{fig:polish}. There is a large
spectral feature at 627 cm$^{-1}$ that appears as an anti-resonance dip in
$\sigma_1$.  This anti-resonance dip is also observed in $\sigma_1$ of the
better-polished surfaces as shown in Fig.~\ref{fig:ac}(a) and
Fig.~\ref{fig:nodal}(a), albeit with less spectral weight, but not in the {\it
ab}-plane conductivity data.  It is due to {\it ab}-plane carriers coupling to
a {\it c}-axis LO phonon \cite{tu02b}, but not caused by direct absorption of
{\it c}-axis TO phonons \cite{reedyk92}.  In the insert of
Fig.~\ref{fig:polish}(a), room temperature reflectance data is given with the
${\rm E}\!\parallel\!{b}$ and ${\rm E}\!\parallel\!{c}$ on this {\it bc} face.
Features associated with {\it c}-axis TO phonons are absent in the E parallel
to {\it b}-axis spectrum. We therefore conclude that there is no significant
contamination by {\it c}-axis phonons in the conductivity data with ${\rm
E}\!\parallel\!{b}$. This indicates that even with this coarser polishing
finish, the planar structure of the Bi2212 is preserved in these polished
surface regions. Still, with coarser polishing there is little evidence of
superfluid response in $\sigma_1$ or $\sigma_2$ similar to what is observed on
better polished {\it ac} and {\it bc} surfaces.

The surprising aspect of our results is the characteristically different
behavior of the optical conductivities in the interface regions of a surface
with the normal along a nodal direction, compared to that of a surface with the
normal along an anti-nodal (gap) direction.  While the superfluid contribution
to optical conductivity can be measured on the $(1\bar{1}0)$-plane, it is
unobservable on the {\it ac} or {\it bc} planes.
%
%
The superconductivity as probed by infrared techniques may behave in a way
similar to what Greene and co-workers have found in their tunneling experiments
\cite{aubin02}.  In the surface region where the node of a {\it d}-wave order
parameter is normal to a reflecting surface like the {\it ac} and {\it bc}
faces of a high-T$_c$ superconductor, Andreev scattering causes strong pair
breaking which leaves a surface region depleted of superfluid. This may explain
why no superfluid response is observed in {\it ac} and {\it bc}-plane
conductivity. The situation is different for the surface region for which the
normal is an anti-nodal direction.  Ideally, there should be no Andreev
scattering because the superconducting pairs do not suffer a change of sign in
the order parameter under a reflection on this interface. This explains why a
superfluid response is observed in the $(1\bar{1}0)$-plane conductivity, albeit
with less magnitude compared to {\it ab}-plane data, and there is a
superconducting gap when tunneling on $(1\bar{1}0)$-plane accompanied by a less
pronounced ZBCP.

%
%
The small jump in $\sigma_1$ and $\sigma_2$ from 100 to 80~K deserves some more
discussion.  A careful study reveals that the increase almost exclusively
occurrs within a few degrees of T$_c$.  We speculate that this jump is related
to the formation of Andreev bound states in the surface region of {\it ac} and
{\it bc} plane as a result of, e.g.~a reduction in scattering rate when bound
states are formed below T$_c$.  The Andreev bound state should deplete the
superfluid to a depth of order the coherence length $\xi_0\sim 100$~\AA .  The
classical skin depth is defined as $\delta = c/\sqrt{2\pi\sigma_1\omega}$,
which is of the order of microns and much greater than the mean-free path, so
that since $\delta \gg \xi_0$ the infrared should still probe the superfluid in
the bulk. This is seemingly at odds with the observation of no further change
in $\sigma_1$ and $\sigma_2$ for $T\ll T_c$. Within the BCS theory, $\xi_0$ can
be defined in terms of the Fermi velocity $v_F$ and the energy gap $\Delta$,
$\xi_0 = \hbar v_F/\pi\Delta$.  However, the energy gap is thought to have a
momentum dependence, thus $\Delta\equiv\Delta_k$.  If $\Delta_k\rightarrow 0$
in the nodal direction, then $\xi_k$ may become quite large, i.e. $\xi_k
\approx \delta$, which would suggest for certain geometries the influence of
the Andreev bound state might extend over a larger region than previously
thought \cite{misra02}.
%
%

%
%
The role of disorder induced by the polishing process also deserves some
further considerations. This kind of problem seems to be reminiscent to the
``two-length scale'' problem in X-ray scattering \cite{andrews86,ryan86}. For
example, in the case of UO$_2$ \cite{watson96} and SrTiO$_3$ \cite{wang98}, it
is found that mechanical processing causes an increase in dislocation density
in the surface region that can be as deep as 500~nm.  However, we do not think
that random disorder can explain the asymmetry we have observed in our optical
measurements nor the asymmetry observed in the tunneling experiments.  Of
course, if the polishing process caused different amounts of damage on two
types of surfaces this could happen, but the data shows there are no large
changes in the normal-state conductivity for the two cases.  This issue of the
extent of the depletion region remains to be understood.


In conclusion, we have observed characteristically different behavior in the
{\it ac} and {\it bc}-plane optical conductivities of optimally-doped Bi2212
single crystals, as compared to the $(1\bar{1}0)$-plane conductivity below
T$_c$. Our observation implies that optical measurements are also sensitive to
the {\it d}-wave nature of the superconducting order parameter in high-T$_c$
cuprates.

%
%
\begin{acknowledgments}
We would like to thank D. N. Basov, A. V. Chubukov, V. J. Emery, P. D. Johnson,
A. Millis, T. Timusk and T. Valla for helpful discussions. This work was
supported by the Department of Energy (DOE) under Contract No.
DE-AC02-98CH10886.  LHG acknowledges support from DOE under Contract No.
DE-FG02-ER9645439, through the Fredrick Seitz Materials Research Laboratory.
\end{acknowledgments}
\vspace*{-0.2cm}

%
%
\bibliography{edge}

\begin{thebibliography}{10}

\bibitem{tsuei94}
C.~C. Tsuei {\it et~al.}, Phys. Rev. Lett {\bf 73},  593  (1994).

\bibitem{wollman95}
D.~A. Wollman, D.~J. {Van Harlingen}, J. Giapintzakis, and D.~M. Ginsberg,
  Phys. Rev. Lett. {\bf 74},  797  (1995).

\bibitem{harlingen95}
D.~J. {Van Harlingen}, Rev. Mod. Phys. {\bf 67},  515  (1995).

\bibitem{covington97}
M. Covington {\it et~al.}, Phys. Rev. Lett. {\bf 79},  277  (1997).

\bibitem{sutherland03}
M. Sutherland {\it et~al.}, Phys. Rev. B {\bf 67},  174520  (2003).

\bibitem{valla99}
T. Valla {\it et~al.}, Science {\bf 285},  2110  (1999).

\bibitem{quijada99}
M.~A. Quijada {\it et~al.}, Phys. Rev. B {\bf 60},  14917  (1999).

\bibitem{tu02a}
J.~J. Tu {\it et~al.}, Phys. Rev. B {\bf 66},  144514  (2002).

\bibitem{geerk88}
J. Geerk, X.~X. Xi, and G. Linker, Z. Phys. B {\bf 73},  329  (1988).

\bibitem{lesueur92}
J. Lesueur, L.~H. Greene, W.~L. Feldmann, and A. Inam, Physica C {\bf 191},
  325  (1992).

\bibitem{wei98}
J.~Y.~T. Wei, N.-C. Yeh, D. Garrigus, and M. Strasik, Phys. Rev. Lett. {\bf
  81},  2542  (1998).

\bibitem{alff98}
L. Alff {\it et~al.}, Phys. Rev. B {\bf 58},  11197  (1998).

\bibitem{krupke99}
R. Krupke and G. Deutscher, Phys. Rev. Lett. {\bf 83},  4634  (1999).

\bibitem{aubin02}
H. Aubin, L.~H. Greene, S. Jian, and D.~G. Hinks, Phys. Rev. Lett. {\bf 89},
  177001  (2002).

\bibitem{hu94}
C.-R. Hu, Phys. Rev. Lett. {\bf 72},  1526  (1994).

\bibitem{tanaka95}
Y. Tanaka and S. Kashiwaya, Phys. Rev. Lett. {\bf 74},  3451  (1995).

\bibitem{fogelstrom97}
M. Fogelstr{\"o}m, D. Rainer, and J.~A. Sauls, Phys. Rev. Lett. {\bf 79},  281
  (1997).

\bibitem{carrington01}
A. Carrington {\it et~al.}, Phys. Rev. Lett. {\bf 86},  1074  (2001).

\bibitem{reedyk88}
M. Reedyk {\it et~al.}, Phys. Rev. B {\bf 38},  11981  (1988).

\bibitem{puchkov96}
A.~V. Puchkov {\it et~al.}, Phys. Rev. Lett. {\bf 77},  3212  (1996).

\bibitem{wang99}
N.~L. Wang, A.~W. McConnell, and B.~P. Clayman, Phys. Rev. B {\bf 59},  576
  (1999).

\bibitem{tajima93}
S. Tajima {\it et~al.}, Phys. Rev. B {\bf 48},  16164  (1993).

\bibitem{homes93}
C.~C. Homes, M. Reedyk, D.~A. Crandles, and T. Timusk, Appl. Opt. {\bf 32},
  2972  (1993).

\bibitem{tu02b}
J.~J. Tu, C.~C. Homes, G.~D. Gu, and M. Strongin, Physica B {\bf 316-317},  324
   (2002).

\bibitem{muller03}
K. A. M{\"u}ller, cond-mat/0306643.

\bibitem{reedyk92}
M. Reedyk and T. Timusk, Phys. Rev. Lett. {\bf 69},  2705  (1992).

\bibitem{misra02}
S. Misra {\it et~al.}, Phys. Rev. B {\bf 66},  100510(R)  (2002).

\bibitem{andrews86}
S.~R. Andrews, J. Phys. C {\bf 19},  3721  (1986).

\bibitem{ryan86}
T.~W. Ryan, R.~J. Nelmes, R.~A. Cowley, and A. Gibaud, Phys. Rev. Lett. {\bf
  56},  2704  (1986).

\bibitem{watson96}
G.~M. Waston {\it et~al.}, Phys. Rev. B {\bf 53},  686  (1996).

\bibitem{wang98}
R. Wang, Y. Zhu, and S.~M. Shapiro, Phys. Rev. Lett. {\bf 80},  2370  (1998).

\end{thebibliography}

\end{document}